\documentclass[fleqn,twoside]{article} 
\usepackage{epsf,multicol,ifthen}
\usepackage{ujp}
\usepackage[cp1251]{inputenc}
\usepackage[english,russian]{babel}
\usepackage{amstext}
\usepackage{amssymb}
\usepackage{cite}
\usepackage {graphicx,amsmath}

\nazva{PLASMA FOR-INJECTOR OF SEPARABLE MATERIAL BASED ON THE
BEAM-PLASMA
DISCHARGE\\ FOR ION-ATOMIC SEPARATION\\ TECHNOLOGIES. CONCEPTION}%

\udk{533.9; 621.039 }

\nazvacol{PLASMA FOR-INJECTOR OF SEPARABLE MATERIAL BASED}%

\avtor{┼.I. SKIBENKO, YU.V. KOVTUN, V.B. YUFEROV}%
\avtorcol{┼.I. SKIBENKO, Yu.V. KOVTUN, V.B. YUFEROV }%

\inst{National Scientific Center ``Kharkiv Institute of Physics
and
Technology''}%
\adr{(1, Akademichna Str., Kharkiv 61108, Ukraine; e-mail:
Ykovtun@kipt.kharkov.ua) }

\begin{document}           
\setcounter{page}{361}%
\maketitl                 
\begin{multicols}{2}
\anot{%
In the paper, the functional definition of a plasma for-injector
of separable material is presented, and the requirements to it are
formulated. The version of a device for the material separation
into elements based on the beam-plasma discharge is under
consideration. The dimensions of a pilot separating device are
determined. The following quantities are estimated: the particle
concentration per unit length of the separating device, effective
length of the beam-plasma interaction (BPI) within the separating
device, dynamics of a plasma density increase for metallic
uranium, and thermal characteristics of a phase transformation
unit. A conclusion was drawn on the expedience and validity of the
development and realization of a plasma for-injector for
separating devices and technologies basing on the beam-plasma
mechanism of formation and heating of a highly ionized plasma.}%

\noindent In the literature [1--3, 9], a possibility to use
magneto-plasma separators (MPS) for the processing of nuclear
waste and spent nuclear fuel was discussed. The idea of the
processing consists in the following: a working material, spent
fuel waste (SFW) or radioactive waste (RAW), is prepared for the
introduction into a phase transformation unit. There the separable
material is transformed from the solid (liquid) state into the
vaporous state and then is introduced into a plasma source chamber
for its ionization. Ions of the plasma formed, being in a magnetic
filed, are selectively heated, which results in changing their
flow path in a magnetic field, the spatial separation of ``hot''
and ``cold'' ions, and their deposition on ion-receiving plates,
from which the deposited elements are removed. In addition, the
plasma ions, from RAW or SFW, can be separated into light and
heavy mass groups (the so-called ``partial separation'') or into
separate elements (``full separation''). For the ``partial
separation'', a problem consists in decreasing the specific dose
of a radioactive part in the stored RAW excluding their complete
processing.\looseness=1

In the operation of the plasma source (injector) based on the
beam-plasma discharge taking place in MPS, one can distinguish
several subsequently running stages. First, it is the stage of
working material preparation in the required phase, namely, the
vaporized state. The second stage intends the introduction
(transport) of a separable material in the vaporized state into
the ionization zone. The third stage is the ionization process
(impact ionization by a linear law). Note that, in the case of the
beam-plasma discharge, the ionization zone can be realized at any
point of the separation volume (track), where the excess density
of neutral particles of the separable matter of an order of
10$^{12}$ ёm$^{ - 3}$ is created. The fourth stage is the
ionization too; but, in this case due to collective processes, the
density increases by a nonlinear (exponential) law. The fifth
stage is specified by the heating of electrons and ions of the
plasma formed in the beam-plasma discharge at the expense of the
electron-cyclotron resonance (ECR) and the ion-cyclotron resonance
(ICR). Thus, the multistage work of the so-called plasma source
can evidence not simply the creation and operation of a plasma
source but the creation and operation of a more complex and more
functional device -- a for-injector for filling the separator
volume by a separable material in the neutral and ionization
states at different time instants.\looseness=1

The main requirements to the plasma source (PS) should be
formulated with regard for the MPS macroparameters (geometry,
dimensions, magnetic field value, diagnostics at the phase of
isotope separation), as well as the plasma microparameters
(density, temperature, their fluctuations, spatial distribution,
ion charge, rates of plasma rotation, plasma stability or
instability). So, the main requirements are:\looseness=1

\noindent -- plasma density at a level of 10$^{10}$ -- 10$^{12}$
ёm$^{ - 3}$;

\noindent -- plasma electron temperature $T_{e} \le 50$~eV;

\noindent -- ion temperature $T_{i} \approx 20$ eV;

\noindent -- plasma stream value $n_{p}v_{p}s$ of an order of
$1\times 10^{19}$ s$^{ - 1}$;

\noindent -- plasma stream velocity $(3 \div 5)\times 10^{4}$
ёm/s;

\noindent -- neutral flux into the separation zone no more than
$3.5\times 10^{18}$ s$^{ - 1}$.

\begin{center} \noindent \epsfxsize=\columnwidth\epsffile{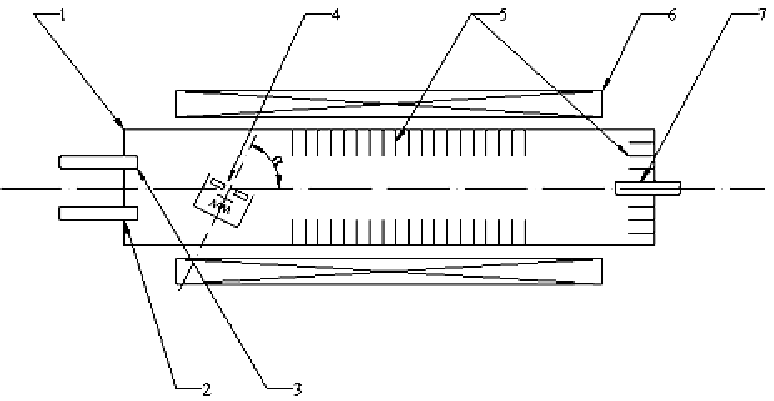}
\end{center}

\vskip-3mm\noindent{\footnotesize Fig. 1. Schematic presentation of the device for material separation into elements}%
\vskip15pt

Furthermore,

\noindent -- ease of placing the ion-receiving plates (electrodes)
for useful product collection;

\noindent -- mode of operation -- stationary or quasistationary;

\noindent -- presence of a natural transverse $E$-field in the
plasma;

\noindent -- providing the useful ion yield onto the ion-receiving
plates under operation conditions without introduction of
additional electric fields and electrodes for their creation;

\noindent -- possibility of reaching and maintaining the more
uniform plasma density distribution across an MPS plasma chamber
during the operating cycle.

\noindent -- possibility of adjusting and controlling the physical
parameters of a plasma source, namely, by changing its process
variables.

The expected parameters of a pilot variant of the separating device can be as
follows:

\noindent -- plasma radius $\sim  0.5$ m;

\noindent -- plasma column length (formation) $\sim 4$ m;

\noindent -- plasma volume 3.14 m$^{3}$;

\noindent -- plasma ion component density $ \ge  10^{12}$ ёm$^{ -
3}$.

Further calculations and estimations will be made with taking
these dimensions and quantities into account.

The schematic diagram of the device separating the material into elements
made on the base of the beam-plasma discharge is presented in Fig. 1 [4].

The device includes vacuum chamber \textit{1} connected with
separable material-feed unit \textit{2} and igniting gas-feed unit
\textit{3}. In its interior, chamber \textit{1} comprises a plasma
source in the form of electron gun \textit{4} and a plasma stream
receiver in the form of plates \textit{5}. The device is provided
with magnetic system \textit{6} embracing chamber \textit{1}. At
the end of vacuum chamber \textit{1}, oppositely to the place of
separable material-feed unit \textit{2}, igniting gas-feed unit
\textit{3} and electron gun \textit{4}, beam collector \textit{7}
is placed. In the device offered, the main physical mechanism~ of~
plasma formation and heating is the collective processes initiated
by the BPI.

The total number of plasma particles $N_{\rm total}^{} $ in the
separator volume can be calculated by the formula
\begin{equation}
\label{eq1} N_{\rm total} = {\int\limits_{S} {n(r)ds(r) = \pi
n_{\max}  r_{0}^{2} {\frac{{\gamma}} {{\gamma + 2}}}}},
\end{equation}

\noindent where $n$ and $ n_{\max}$ are the plasma density and its
maximum, respectively,$ r_{0}$ is the maximum radius of a plasma
formation, and \textit{$\gamma $} is the index characterizing the
profile, i.e. the type of a spatial plasma density distribution.
At $\gamma  = 2$, the plasma density distribution
${\frac{{n(r)}}{{n_{\max}} } } = 1 - \left( {{\frac{{r}}{{r_{0}}}
}} \right)^{\gamma} $is parabolic; at $\gamma   \ge  3$, it is
close to the equilibrium one; and, at $\gamma  < 2$, it is tapered
to the periphery.

In [5], it is shown that, in the plasma formed as a result of the
beam-plasma interaction development, the exponent $\gamma $ as a
function of charge parameters changes from 0.3 to 10. This
evidences that the total number of particles in the discharge (in
the separator chamber) can differ by a factor of 5--6 and more
depending on the plasma profile. The maximum value of $N_{\rm
total} $ can be reached at $\gamma   \sim 10$, i.e. in the case of
a uniform plasma density distribution across the plasma section.
For the above-mentioned separator dimensions, the total number of
particles per unit length of the separator is $0.39\times 10^{19}$
particle/m and $0.65\times 10^{19}$ particle/m for the parabolic
plasma density distribution ($\gamma  = 2$) and for the uniform
distribution ($\gamma  = 10$), respectively (see Fig.~2).

As is known [6], the beam-plasma discharge formation is critical to the
interaction length. This is explained by the fact that there is the least
length, at which the excitation of oscillations to the noticeable amplitude
is possible. The minimum length, at which the beam dissipates its energy for
the excitation of plasma oscillations, can be estimated by the formula [6]
\begin{equation}
\label{eq2}
L \approx {\frac{{v_{0}}} {{\gamma}} } \approx 10^{ - 8}{\frac{{E_{e}
}}{{j}}}\sqrt {n_{p}},
\end{equation}

\noindent where $\gamma $ is the oscillation amplitude rise
increment; $v_{0}$ is the directed electron velocity in the beam;
$E_{e}$ is the beam electron energy in eV; $j$ is the beam current
density in └/ёm$^{2}$; and $n_{p}$ is the plasma concentration in
ёm$^{ - 3}$.

The oscillation amplitude increases with the length of the
electron beam-plasma interaction and, at some lengths, can reach
the value sufficient for the additional gas ionization. Beginning
from this length, a plasma-beam discharge occurs. Figure 3 shows
the dependences

\begin{center} \noindent \epsfxsize=\columnwidth\epsffile{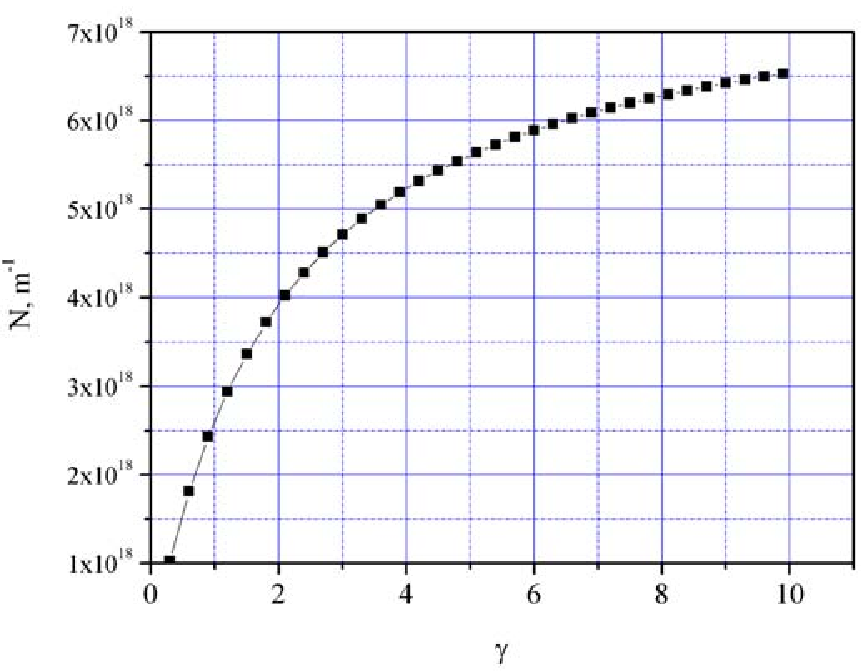}
\end{center}

\vskip-3mm\noindent{\footnotesize Fig. 2. Particle number per unit length of the separating device as a function of the index characterizing the plasma density radial distribution profile}%
\vskip15pt

\noindent of the effective BPI length on the electron beam current
density in the range of $j = 0 \div 50$ └/ёm$^{2} $at different
values of its energy (1.5 and 15 keV) for a plasma density of
10$^{12}$ ёm$^{ - 3}$. It is seen that the calculated values of
the effective BPI length are ranging from fractions of 1 cm to
tens and hundreds of centimeters. For the really obtained electron
beam current (10 -- 20 A) at transverse dimensions 2$R \approx 1$
cm, we have the current density $j \approx  10 \div 25$ └/ёm$^{2}$
and, respectively, the interaction length $L \approx 10 \div 20$
ёm. At higher current density values, for example $ \le $ 100
└/ёm$^{2}$, the required interaction lengths are at the same level
or a slightly decreased one.

Taking into account the literature data [7] on the ratio
$r_{p}/r_{b}\approx 5\div 10$ ($r_{p}$ is the plasma radius,
$r_{b} $ is the electron beam radius), we obtain that, in the
given project at $r_{p} = 50$ ёm, the value of $r_{b} $ should be
not less than 10--15 cm. For this, it is necessary to obtain
electron beams of a larger cross-section (of a large aperture)
and, respectively, cathodes with a larger emitting surface
200--400 cm$^{2}$. Such an emitting surface can be obtained either
in the form of a synthesized polyelement (mosaic) block or on the
plasma cathode base [8]. At an electron beam current density of
1--2 A/cm$^{2}$, the power supplied for its formation is of an
order of 1 MW. At the same time, the power of a RF-oscillator used
for the realization of the project ARCHIMED is 8 MW [2].

The processes running due to the electron beam-gas (vapor) target
interaction and leading to the dense plasma formation can be divided into
two characteristic stages. At the first stage, the electron beam with density
$n_{e}$ and velocity $v_{e}$ interacts with the neutral gas (vapor)

\begin{center} \noindent \epsfxsize=\columnwidth\epsffile{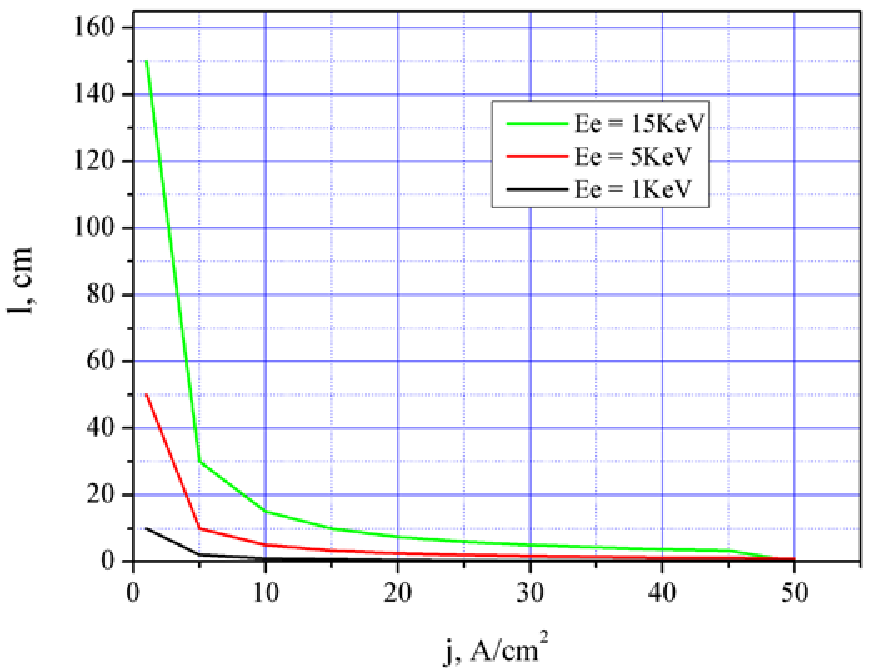}
\end{center}

\vskip-3mm\noindent{\footnotesize Fig. 3. Effective beam-plasma
interaction length as a function of the electron beam current
density for different energy values at a plasma density $n_p =
10^{12}$ ёm$^{-3}$} \vskip15pt

\noindent of a density $n_{0}$, and, due to the impact ionization,
a primary plasma with density $n_{p}$ less or equal to the beam
density is formed. The process of plasma formation at this
instance of time is described by the equation
\begin{equation}
\label{eq3}
{\mathop {n}\nolimits_{p}}  = {\mathop {n}\nolimits_{e}} {\mathop
{v}\nolimits_{e}} {\mathop {\sigma} \nolimits_{e}} {\mathop {n}\nolimits_{0}
}{\mathop {\tau} \nolimits_{}},
\end{equation}
where $\sigma _{e}$ is the ionization cross-section of neutral particles
with electrons, and $\tau $ is the plasma lifetime.

The degree of ionization is at a level of $\sim$$10^{ - 2}${\%},
and the temperature is lower than the first ionization potential
($T_{e} < E_{i})$. The beginning of the second potential is
characterized by the excitation of sufficiently powerful rf
oscillations at the electron-cyclotron frequency [9]. The plasma
electrons, being accelerated in the fields of these oscillations,
reach the energy necessary for the further neutral gas ionization,
and the increase of the plasma density is avalanche-like, the
degree of ionization increases to 100{\%}, and the density
increases by the exponential law
\begin{equation}
\label{eq4}
dn_{p} = {\left\langle {\sigma _{e} v_{e}}  \right\rangle} n_{0} n_{p} dt ,
\end{equation}
where $n_{p}$ is the current plasma density.

Here, it is assumed that the time of the plasma electron
acceleration is much less than the time of the plasma density
increase, and the plasma losses at $n\ll n_{0}$ are negligibly
small. Such a statement is valid only for the instant of time
$t\ll \tau $. Later on, the plasma density increase will be
determined to a greater extent by the balance of the exponential
density growth according to

\end{multicols}
\begin{multicols}{2}
\begin{center} \noindent \epsfxsize=\columnwidth\epsffile{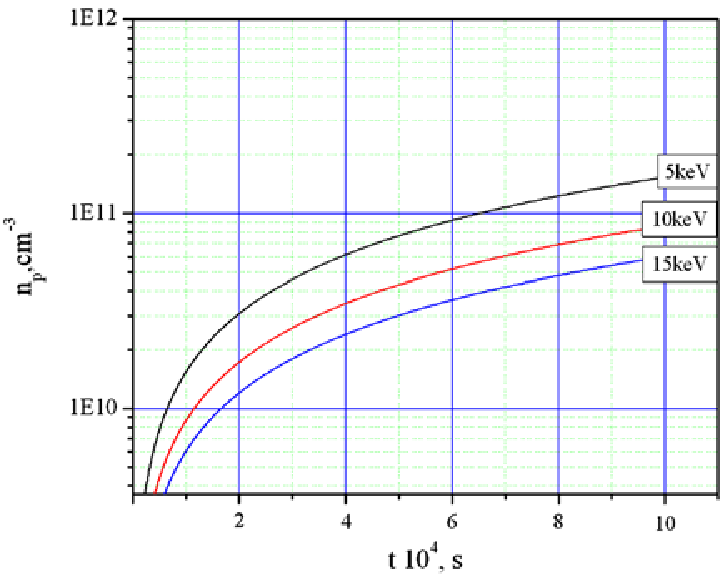}

{\bf a}

\noindent \epsfxsize=\columnwidth\epsffile{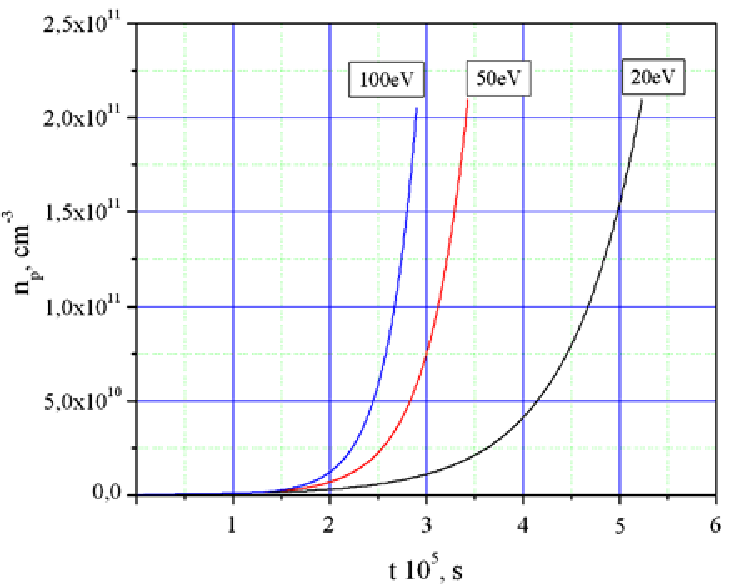}

{\bf b}
\end{center}
\end{multicols}

\vskip-3mm\noindent{\footnotesize Fig. 4. Plasma density increase
dynamics calculated for metallic uranium: {\it р} -- linear stage
of discharge; {\it b} -- exponential stage of discharge}%
\vskip25pt
\begin{multicols}{2}

\noindent Eq. (\ref{eq4}) and by the plasma loss. Among the
processes leading to the plasma loss, the processes of plasma
recombination and diffusion across the magnetic field and its
runaway into the magnetic plug are dominant.

Figure 4 presents the calculated time dependences of the plasma
density for the linear (\textit{a}) and exponential (\textit{b})
growth in the case of metallic uranium [10, 11] during its
evaporation or corpuscular-plasma sputtering in the discharge
zone. In addition, to evaporate a material, one can use the
electron-beam evaporation, laser evaporation, flash evaporation
[12--14]. An important feature of the electron-beam evaporation is
that it can be realized in variants with or without crucible [13].

An important characteristic of the whole plasma source and its components is
the energy expenditure for the realization of phase transformations of a
separable material and its ionization. To determine the plasma source
efficiency, we estimated the energy expenditures, the required working
material mass consumption at a constant plasma stream, and the specific rate
of material evaporation in the phase transformation unit for groups of
metals, being of interest for imitation-separation experiments, as well as
for the industrial and semiindustrial processing of RAW and
SFW.

For the metal under consideration at a constant plasma stream at a
level of $4.7\times  10^{21} $ particle/s, the values of the
required material mass consumption $m$, specific rate of material
evaporation $a_{V}$, evaporation temperature $T_{V}$, and melting
temperature $T_{\rm mel}$ are given in Table 1.

The calculated values of the molar quantity of heat for the evaporation of
selected materials (for UO$_{2}$, it is the molar melting heat), as well as
the molar energy expenditures by electron-beam evaporation of Zr and U [15]
are given in Table 2.

The use of BPI makes it possible to obtain the plasma with given
parameters ($n_{e}\sim  n_{i}  \ge  10^{12}$ ёm$^{ - 3}$; $T_{e}
 \ge  100$ eV; $T_{i}  \ge  300$ eV) in large volumes
($\sim 1$--10 m$^{3})$ of separating devices. The operation mode
of the beam-plasma discharge can be pulsed (100--600 $\mu $s),
quasistationary (1--1000 $\mu $s), or stationary (1--10 s and
more).

The calculations show that the effective length of the primary
electron beam deceleration is from tens of centimeters to several
meters, which  corresponds, in principle, to linear dimensions of
separating devices and

\vskip3mm
\noindent{\footnotesize{\bf%
 T a b l e ~1}\vskip1mm \tabcolsep9.2pt

\noindent\begin{tabular}{c c c c c}
 \hline \multicolumn{1}{c}
{\rule{0pt}{9pt}} & \multicolumn{1}{|c}{$\dot{m}$, g/s}&
\multicolumn{1}{|c}{$a_{V}$, g/(cm$^2 \times$s)}&
\multicolumn{1}{|c}{$T_V$, K}& \multicolumn{1}{|c}{$T_{\rm mel} $, K}\\%
\hline%
\rule{0pt}{9pt}Zr & 0.719& $1\times 10^{-2}$&  3189&    2133\\%
Bi& 1.646& $2.5\times 10^{-2}$& 1153& 544\\%
Pb&  1.632& $1.8\times 10^{-2}$& 1254& 600.65\\%
U& 1.875& $1.7\times 10^{-2}$& 2781& 1408\\%
UO$_2$& 2.127&$1.7\times 10^{-2}$& 2800& 3123\\%
\hline
\end{tabular}
}

\vskip3mm
\noindent{\footnotesize{\bf%
 T a b l e ~2 }\vskip1mm \tabcolsep15.1pt

\noindent\begin{tabular}{c c c }
 \hline \multicolumn{1}{c}
{\rule{0pt}{9pt}} & \multicolumn{1}{|c}{$Q \times 10^{-5}$,
J/mole}
& \multicolumn{1}{|c}{$E\times 10 ^{-7}$, J/mole}\\%
\hline%
\rule{0pt}{9pt}Zr&7.63&2.16\\%
Bi&2.3~~&--\\%
Pb&2.33&--\\%
U&6.46&1.8~~\\%
UO$_2$&3.78&--\\%
 \hline
\end{tabular}
}

\noindent keeps within the range of separation of elements or
their isotopes.

The major advantage of the plasma formation method offered is the fact that the
electron beam from an external source (electron gun) in vacuum and
a longitudinal magnetic field propagates almost without losses at any
distances within the limits of the above-mentioned values. That is, the plasma can
be formed at any points of the transport track having the length of several
meters, namely in the separation zone.

Another substantial merit of this method, as compared with the other plasma
formation methods, is the evidence of that, in the mass composition of the
plasma formed, only the particles (ions, neutrals) of the introduced working
material are presented, and the plasma is not contaminated with particles
from materials of electrodes, diaphragms, etc.

Under conditions of the beam-plasma discharge, it is possible to
reach the 100-{\%} neutral burn-up that is confirmed with the
confidence in earlier experiments~[16].

The realization of the beam-plasma discharge allows one to use different methods
for the introduction of a working material, practically, at any point (into a region) of
the injection track.

So, the beam-plasma mechanism of formation of a dense hot
high-ionized plasma gives promise for the development and the
realization of a plasma for-injector on its base, as well as its
use in separating devices and technologies.

\rezume{%
 ╧╦└╟╠╬┬╚╔~~~~ ╘╬╨▓═╞┼╩╥╬╨~~~~ ╨╬╟─▓╦▐┬└═╬п \\╨┼╫╬┬╚═╚~ ═└ ~╬╤═╬┬▓
╧╙╫╩╬┬╬-╧╦└╟╠╬┬╬├╬ \\╨╬╟╨▀─╙ ─╦▀ ▓╬══╬-└╥╬╠═╚╒ ╤┼╧└╨└╓▓╔═╚╒
\\╥┼╒═╬╦╬├▓╔. ╩╬═╓┼╧╓▓▀}{к.▓. ╤ъ│схэъю, ▐.┬. ╩ютЄєэ, ┬.┴. ▐ЇхЁют}
{═ртхфхэю ЇєэъЎ│юэры№эх тшчэрўхээ  яырчьютюую ЇюЁ-│эцхъЄюЁр
Ёючф│ы■трэю┐ Ёхўютшэш │ ёЇюЁьєы№ютрэю тшьюуш фю э№юую. ╨ючуы эєЄю
трЁ│рэЄ яЁшёЄЁю■ фы  Ёючф│ыхээ  Ёхўютшэш эр хыхьхэЄш эр юёэют│
яєўъютю-яырчьютюую ЁючЁ фє. ┬шчэрўхэю Ёючь│Ёш эря│тяЁюьшёыютюую
трЁ│рэЄр ёхярЁрЎ│щэюую яЁшёЄЁю■. ╬Ў│эхэю ъюэЎхэЄЁрЎ│■ ўрёЄшэюъ эр
юфшэшЎ■ фютцшэш ёхярЁрЎ│щэюую яЁшёЄЁю■, хЇхъЄштэє фютцшэє
яєўъютю-яырчьютю┐ тчр║ьюф│┐ т ьхцрї ёхярЁрЎ│щэюую яЁшёЄЁю■,
фшэрь│ъє ЁюёЄє ∙│ы№эюёЄ│ яырчьш ьхЄры│ўэюую єЁрэє, Єхяыют│
їрЁръЄхЁшёЄшъш сыюър Їрчютшї яхЁхЄтюЁхэ№. ╟Ёюсыхэю тшёэютюъ яЁю
фюЎ│ы№э│ёЄ№ │ юс┤ЁєэЄютрэ│ёЄ№ ЁючЁюсъш │ Ёхры│чрЎ│┐ яырчьютюую
ЇюЁ│эцхъЄюЁр фы  ёхярЁрЎ│щэшї яЁшёЄЁю┐т │ Єхїэюыюу│щ эр юёэют│
яєўъютю-яырчьютюую ьхїрэ│чьє єЄтюЁхээ  │ эруЁ│трээ 
тшёюъю│юэ│чютрэю┐ яырчьш.}

\end{multicols}
\end{document}